\documentclass[a4paper,11pt]{article}
\pdfoutput=1 

\usepackage{jcappub} 
\usepackage{color}

\usepackage[T1]{fontenc} 

\newcommand{\be}{\begin{eqnarray}}
\newcommand{\ee}{\end{eqnarray}}

\title{Failure of the stochastic approach to inflation beyond slow-roll}

\author[a]{Diego Cruces,}
\author[a,b]{Cristiano Germani,}
\author[c]{Tomislav Prokopec}

\affiliation[a]{Institut de Ciencies del Cosmos (ICCUB), Universitat de Barcelona, Mart\'i i Franqu\`es 1,
E08028 Barcelona, Spain}
\affiliation[b]{Departement de F\'isica Qu\`antica i Astrofisica, Universitat de Barcelona, Mart\'i i Franqu\`es 1, 08028 Barcelona, Spain}
\affiliation[c]{Institute for Theoretical Physics, Spinoza Institute and EMME$\Phi$, Utrecht University,\\
Postbus 80.195, 3508 TD Utrecht, The Netherlands}

\emailAdd{germani@icc.ub.edu}
\emailAdd{dcrucema7@alumnes.ub.edu}
\emailAdd{t.prokopec@uu.nl}

\abstract{After giving a pedagogical review we clarify that the stochastic approach to inflation is generically reliable only at zeroth order in the (geometrical) slow-roll parameter $\epsilon_1$ if and only if $\epsilon_2^2\ll 6/\epsilon_1$, with the notable exception of slow-roll. This is due to the failure of the stochastic $\Delta N$ formalism in its standard formulation. However, by keeping the formalism in its regime of validity, we showed that, in ultra-slow-roll, the stochastic approach to inflation reproduces the power spectrum calculated from the linear theory approach.
}

\begin{document}
\maketitle
\flushbottom

\section{Introduction}

After a renewed interest of primordial black holes (PBH) as dark matter (DM) candidates \cite{pbhs},  ultra-slow-roll (USR) \cite{Kinney:2005vj,Martin:2012pe} has attracted a great deal of attention as a transient phase of single-field inflation. Indeed, while an initial slow-roll (SR) phase of inflation predicts the observed cosmic microwave background radiation (CMB), a subsequent USR trajectory of the inflaton can generate a peak in the power spectrum of curvature perturbations, at scales smaller than those of the CMB \cite{tomi}. Although the amplitude of this peak can be several order of magnitudes larger than the one of the CMB's power spectrum, it can nevertheless still be small enough not to invalidate linear perturbation theory. Gravitational collapse into PBHs requires the curvature perturbations to be already in the non-linear regime at super-horizon scales, when the above peak is generated. Assuming a Gaussian distribution for the amplitude of cosmological perturbations, rare non-linear fluctuations are statistically generated and, if certain conditions are met \cite{musco}, they would collapse into PBHs at horizon re-entry. The abundances of these PBHs is clearly related to the hight of the generated peak and therefore a minimal value of it is required in order to have enough DM at later times \cite{musco}.

One might question whether the Gaussian assumption is justified \cite{toni}. In case of a smooth transition between the SR and USR phase it has been proven that it is \cite{sasaki}. The second question that might come naturally into mind is whether or not the effects of quantum diffusion arising from non-linear quantum effects could be large enough to spoil the predictions of the linear quantum theory calculated in an USR inflationary phase. The suspicion
that this might happen comes from the fact that in USR the inflaton potential is exactly flat. In this case, the inflaton velocity is completely dominated by Hubble friction and, as time passes, it becomes exponentially suppressed. 
Therefore, additional quantum kicks generated by non-linear interactions in the effective Lagrangian of the inflaton,
might eventually become significant, thus changing appreciably the prediction of the power spectrum obtained 
by the linear analysis.  

A natural framework to study this question is stochastic inflation \cite{Starobinsky3}. The first paper asking whether or not quantum diffusion would be important in the context of PBHs formation was \cite{vennin}. Although from the study of \cite{vennin} one would be tempted to extrapolate that quantum diffusion effects are important in USR, this would not necessarely be true as the perturbations in SR and USR behave quite differently. A subsequent paper \cite{riotto2} {\it assumed} that, at order zero in the slow-roll parameters, the power spectrum of curvature perturbations is exactly the same as the one calculated by linear perturbation theory. We shall show that is correct since, without any slow-roll terms, the equations governing the inflaton in the stochastic approach are linear. However, 
this result is at odds with \cite{bellido}.  In the latter paper, the power spectrum of curvature perturbations in Fourier space, has been calculated as if curvature perturbations would be constant at super-horizon scales. However, this is unfortunately not correct for the cases of USR (and slower) trajectories. Using instead the fact that scalar field fluctuations {\it are} constant at super-horizon scales and that at the zeroth order in the slow-roll parameters stochastic inflation is linear, allows us to reconstruct the power spectrum of the curvature perturbation with the reassuring result that it matches the one from the linear perturbation analysis. 

Having settled that, at zeroth order in slow roll parameters, the power spectrum can be obtained by studying the free (quantum) diffusion. However, one may still wonder whether slow-roll and nonlinear corrections give us more information. Unfortunately, the stochastic framework in USR and in constant-roll is not valid at the first sub-leading (linear) 
order in slow-roll parameters and thus other quantum field theory methods should be used. This observation constitutes one of the main results of our paper.

Throughout this work we use natural units in which the speed of light $c=1$, the reduced Planck constant $\hbar=1$ and 
we often express the Newton constant $G$ in terms of the reduced Planck mass, 
$M_{\rm P}=1/\sqrt{8\pi G}\simeq 2.45\times 10^{18}~{\rm GeV}$.

\section{Review of stochastic inflation}

For inflation one usually considers one (or several) scalar field(s) $\phi$ called the ``inflaton'' which dominates the energy density of matter in the early universe. Here we will however focus on the single field case. The action of such a system is then
\begin{align} \label{Introduction}
S=\frac{1}{16\pi G}\int d^4x\sqrt{-g}R +\int d^4x\sqrt{-g}
\left[
 -\frac12g^{\mu\nu}\partial_\mu\phi\partial_\nu\phi-V(\phi)
\right]
\,.
\end{align}
On a FRW background $\bar{g}_{\mu\nu}$ is given by
\begin{equation}
\bar g_{\mu\nu} = {\rm diag}\Big(-1,a^2(t),a^2(t),a^2(t)\Big)
\,,
\label{cosmological metric}
\end{equation}
while the equation of motion for the homogeneous scalar field $\phi_0(t)\equiv \langle\phi(t,\vec x)\rangle$ is 
\begin{equation}
\left(\frac{\partial^2}{\partial t^2}+3H\frac{\partial}{\partial t}\right)\phi_0(t)+V'(\phi_0)=0
\,,
 \label{eq1}
\end{equation}
with $a=a(t)$ the scale factor and $H=H(t)$ is the Hubble parameter
that obeys the constraint and dynamics imposed by gravity 
\begin{equation}
H^2 = \frac{\rho_\phi}{3M_{\rm P}^2}
\,,\qquad 
\dot H = -\frac{\rho_\phi+P_\phi}{2M_{\rm P}^2}
\,,
\label{Friedman equations}
\end{equation}
where $\rho_\phi = \frac{\dot\phi_0^2}{2}+V(\phi_0)$ 
and $P_\phi=\frac{\dot\phi_0^2}{2}-V(\phi_0)$ are the energy density and pressure
of the spatially homogeneous part of the inflaton fluid, respectively.

It is convenient to present here equations \eqref{eq1} and \eqref{Friedman equations} in conformal time $d\tau=dt/a$ since they will be useful throughout this work,
\begin{align} 
\label{eq1conformal}
\left(\frac{\partial^2}{\partial \tau^2}+2{ \cal H}\frac{\partial}{\partial \tau}\right)\phi_0(\tau)+a^2V'(\phi_0)=0
\,,\\ \label{Friedmanconformal}
{\cal H}^2 = \frac{\rho_\phi}{3M_{\rm P}^2}a^2
\,,\qquad 
{\cal H}'= -\frac{\rho_\phi+3P_\phi}{6M_{\rm P}^2}a^2
\,,
\end{align}
where ${\cal H}'=\frac{\partial {\cal H}}{\partial\tau}$ and where we made use of, 
\begin{equation} \label{confchanges}
\frac{\partial}{\partial t}=\frac{1}{a}\frac{\partial}{\partial \tau}
\,,\quad 
\frac{\partial^2}{\partial t^2}=
 \frac{1}{a^2}\left(\frac{\partial^2}{\partial\tau^2}-{\cal H}\frac{\partial}{\partial\tau}\right)
\,,\quad
{\cal H}=\frac{a'}{a}=Ha
\,.
\end{equation}
Since inflation occurs when the field rolls very slowly compared with the expansion of the universe \cite{LindeInflation}, the usual way of studying \eqref{eq1} is the SR approximation \cite{Liddle}, where one neglects the acceleration of the field
($d^2 \phi_0/dt^2$), so that the scalar is pulled by a nearly flat potential. In the USR case instead the potential is essentially flat
$\left(d V(\phi_0)/d\phi_0=0=d^2 V(\phi_0)/d\phi_0^2\right)$
 so the acceleration of the field becomes important.  

Usually, in the standard description of inflation, one supposes the homogeneous part of the field to follow a classical trajectory whereas the small deviations from homogeneity are treated quantum mechanically \cite{mukhanov}. However, one could expect a modification in the classical trajectory of the field due to quantum effects \cite{Vilenkin, Salopek}. In order to incorporate these quantum effects to the classical trajectory, the stochastic formalism was born \cite{Starobinsky3, Kandrup, Salopek2}.

The basic idea of stochastic inflation is to reduce a suitably coarse-grained evolution of the full quantum inflaton field  dynamics to a much simpler, 
but almost equivalent, stochastic problem. 
This is done by splitting the inflaton into a quantum short-scales part, in which the field is fully quantum but for which perturbative
methods apply, and a stochastic large-scale part, in which the field is influenced by the quantum sector by receiving kicks from an approximately Markovian stochastic force.

Before introducing the Langevin equation, which is the dynamical equation of the stochastic infrared field, 
it is useful to remind the reader about the linear quantum field theory computation in both SR and USR.

\subsection{The Mukhanov-Sasaki equation for linear perturbations}

We can now consider small fluctuations to the inflationary trajectory. Since the stochastic formalism studies only the quantum deviations of the inflaton with respect to its classical trajectory, we will consider the spatially flat gauge where the scalar field can be split as a background value $\phi_0$ and a fluctuation $\delta\phi$, whereas the scalar metric fluctuations are set to zero.

In this gauge~\footnote{One can show~\cite{Mukhanov:1990me} that, up to boundary terms, Eq.~(\ref{quantumaction})
 is the gauge invariant action for the Mukhanov variable~$v=a[\delta\phi+(\dot\phi_0/H)]\psi$, where $\psi$ Is the spatial gravitational
potential perturbation.} 
one has~\cite{Mukhanov:1990me}
\begin{align} 
\label{quantumaction}
S=\int{\cal L}d\tau d^3x=\frac{1}{2}\int\left(v'^2+v\Delta v+\frac{z''}{z}v^2\right)d\tau d^2x\,,
\end{align}
where $v=a\delta \phi$, $z=(a \phi_0')/{\cal H}$ and $\Delta=\sum_{i}\partial_i^2$ is the spatial Laplacian operator.

In terms of slow-roll parameters
\begin{align}
z = a\frac{\phi_0'}{{\cal H}}=a\sqrt{2\epsilon_1}M_{\rm P}\,,\\
\frac{z''}{z} = {\cal H}^2\left(2-\epsilon_1 + \frac32\epsilon_2 
 -\frac12\epsilon_1\epsilon_2 + \frac14\epsilon_2^2+\frac12\epsilon_2\epsilon_3\right)
\,,
\label{introducing z}
\end{align}
where we have used the geometric definitions 

\begin{equation}
 \epsilon_1 = \frac{\phi_0'^2}{2{\cal H}^2 M_{\rm P}^2} = -\frac{\partial}{\partial N}\ln H
\,,\qquad 
   \epsilon_2 = \frac{\epsilon_1'}{\epsilon_1 {\cal H}} = \frac{\partial}{\partial N}\ln(\epsilon_1)
\,,\qquad 
   \epsilon_3 = \frac{\epsilon_2'}{\epsilon_2 {\cal H}} = \frac{\partial}{\partial N}\ln(\epsilon_2)
   \,.
\label{geometric slow roll parameters: stochastic def}
\end{equation}
and the number of e-folds $dN=Hdt={\cal H}d\tau$ has been used as a time variable.

In SR $\epsilon_{i+1}{,}\epsilon_i\ll 1$ and therefore
\be
\frac{\left(z^{(SR)}\right)''}{z^{(SR)}} = {\cal H}^2\left(2-\epsilon_1 + \frac32 \epsilon_2 + {\cal O}(\epsilon_i^2)\right)\ .
\ee
The situation is substantially different in USR. It is easy to see that 
\begin{align}
\epsilon_1^{(USR)}=\frac{\left(\frac{\partial \phi}{\partial N}\right)^2}{2M_P^2}\propto \frac{e^{-6N}}{H^2}\,.
\end{align}
This result allows us to compute $\epsilon^{(USR)}$'s of higher order: 
\begin{eqnarray} \nonumber
\epsilon_2^{(USR)} & = & \frac{d}{dN} \ln(\epsilon_1^{(USR)}) =\frac{1}{\epsilon_1^{(USR)}}\frac{d\epsilon_1^{(USR)}}{dN}=-6+2\epsilon_1^{(USR)}\,,\\ \nonumber
\epsilon_3^{(USR)} & = & \frac{1}{\epsilon_2^{(USR)}}\frac{d\epsilon_2^{(USR)}}{dN} =2\epsilon_1^{(USR)} \,,\\ \nonumber
\epsilon_4^{(USR)} & = & \frac{1}{\epsilon_3^{(USR)}}\frac{d\epsilon_3^{(USR)}}{dN} =\epsilon_2^{(USR)} \,,\\ \label{epsilons}
& \vdots & \,.
\end{eqnarray}
Equations \eqref{epsilons} can be written in a compact way:
\begin{eqnarray} \nonumber
\epsilon_n^{(USR)}=-6+2\epsilon_1^{(USR)}\,, & \qquad \text{when n even}\,, \\ \label{epsilons2}
\epsilon_n^{(USR)}=2\epsilon_1^{(USR)}\,, & \qquad \text{when n>1 and odd} \,.
\end{eqnarray}
In this case we finally have 
\begin{align} \label{zzUSR}
\frac{\left(z^{(USR)}\right)''}{z^{(USR)}}={\cal H}^2\left(2-7\epsilon_1^{(USR)}+{\cal O}((\epsilon_1^{(USR)})^2)\right)\,.
\end{align}
We immediately see then that $\frac{\left(z^{(USR)}\right)''}{z^{(USR)}}$ and $\frac{\left(z^{(SR)}\right)''}{z^{(SR)}}$ differ already at order $\epsilon_1$. 
\subsubsection{Solution of the Mukhanov Sasaki equation}

In order to integrate out the short wavelength modes, we will need the solution of the Mukhanov-Sasaki equation (MS), obtained by varying the action \eqref{quantumaction} with respect to $v$. By defining the new rescaled variable $v\equiv (-\tau)^{1/2}s$, the MS equation in Fourier space becomes
\begin{align} \label{svar}
\left\lbrace\tau^2 \frac{\partial^2}{\partial \tau^2}+\tau \frac{\partial}{\partial \tau}+ \left[k^2\tau^2-\left(\frac{1}{4}+\frac{z''}{z}\tau^2\right)\right]\right\rbrace s(\tau,k)=0\,.
\end{align}
During inflation $\nu^2=\frac{1}{4}+\frac{z''}{z}\tau^2$ is an adiabatic function of time~\footnote{Adiabaticity is respected
both during SR and USR, but may be broken at the transition between the two. Even though potentially interesting, 
in this work we do not consider consequences of the possible break down of adiabaticity during the transition.}. In other words its rate of change is much smaller than the Hubble rate. Thus, approximating it as a constant, the solutions can be written 
in terms of Hankel functions of the first and second kind $H_{\nu}^{(1)}(-k\tau)$ and $H_{\nu}^{(2)}(-k\tau)$, i.e.
\begin{align} \label{solution}
s(\tau,k)=(-\tau)^{-1/2}v(\tau,k)=C_1(k)H_{\nu}^{(1)}(-k\tau)+C_2(k)H_{\nu}^{(2)}(-k\tau)\ .
\end{align}
The variation of $\nu$ in time would introduce a decaying mode that we discard.

In order to set the constants $C_1(k)$ and $C_2(k)$ one uses the Bunch-Davies vacuum by noticing that at very short scales the system behave like an harmonic oscillator and therefore can be easily quantized. By doing that, one obtains that the magnitude of the quantum fluctuations in spatial Fourier space $\delta \phi_{\textbf{k}}(\tau)$ are given in terms of the mode functions
\begin{align} \label{SOL}
\varphi_k(\tau)=\frac{v_k(\tau)}{a}=\frac{\sqrt{-\tau}}{a}\sqrt{\frac{\pi}{4}}H^{(2)}_{\nu}(-k\tau)\,,
\end{align}
as
\begin{align}
 \label{eq.0.2}
\delta\phi_{\textbf{k}}(\tau)=a_{\textbf{k}}\varphi(\tau, k)+a_{-\textbf{k}}^{\dagger}\varphi^{*}(\tau, k)
\,,
\end{align}
with $a_{\textbf{k}}$ and $a_{\textbf{k}}^{\dagger}$ being the usual quantum creation and annihilation operators.

\subsection{The Langevin equation}

The stochastic formalism uses the separate universes approach \cite{Wands} which consists into two separate assumptions: (a) at super-Hubble scales spatial gradients can be neglected (b) the evolution of the gauge invariant scalar perturbations is well approximated by a Klein-Gordon equation in a local FRW space-time. The scalar field equation in this separate universe approach is then \eqref{eq1}:
\begin{align} \label{new1}
\left(\frac{\partial^2 }{\partial t^2}+3H\frac{\partial }{\partial t}\right)\phi_{IR}+V'(\phi_{IR})=0
\,.
\end{align}
Note that in the equation above we have denoted the inflaton field as $\phi_{IR}$ because we are talking about the $\phi$ modes 
with comoving wavelength $\lambda_c\sim 1/k$ larger than the Hubble radius.

In the stochastic approach the IR modes receive stochastic kicks from short-wavelength modes (UV modes), where the separate universe approach does not hold. As the short scale modes are in the perturbative regime, 
to leading (linear) approximation in the UV modes, schematically one ought to solve the following equation,
\begin{align} \label{new2}
KG(\phi_{IR})+MS(\phi_{UV})=0
\,,
\end{align}
where $KG$ stands for the operator in~\eqref{eq1}, and $MS$ the Mukhanov-Sasaki (MS) operator discussed earlier, 
$MS\equiv (1/a^3)[\partial_\tau^2-\Delta-(z''/z)]a$. 

Equation~\eqref{new2} can be written in conformal time as follows,
\begin{align} \label{new30}
\left(\frac{\partial^2 }{\partial \tau^2}+2{\cal H}\frac{\partial }{\partial \tau}\right)\phi_{IR}+a^2V'(\phi_{IR})+\frac{1}{a}\left(v''-\Delta v-\frac{z''}{z}v\right)=0
\,,
\end{align}
where here $v=a\delta\phi_{UV}$ and where $\frac{z''}{z}$ is defined in \eqref{introducing z}. 

We can now modify the second parenthesis in \eqref{new30} in order to solve the MS equation for $\delta\phi_{UV}$ instead of for $v$. The calculation is straightforward and the result is:
\begin{align} \nonumber
\left(\frac{\partial^2 }{\partial \tau^2}+2{\cal H}\frac{\partial }{\partial \tau}\right)\phi_{IR}
+{a^2}V'(\phi_{IR})+\\\label{new3}
\left[\frac{\partial^2}{\partial \tau^2}+2{\cal H}\frac{\partial }{\partial \tau} +{\cal H}^2\left(-\frac{\Delta}{(Ha)^2}-\frac32\epsilon_2+\frac{1}{2}\epsilon_1\epsilon_2-\frac14\epsilon_2^2-\frac{1}{2}\epsilon_2\epsilon_3\right)\right]\delta\phi_{UV}=0
\,.
\end{align}

It has been shown in \cite{Vennin1, Finelli:2008zg}, that the correct stochastic time is $N$\footnote{Note however that in Ref.~\cite{Vennin1},
as well as in related references, the authors neglect the second time derivative of the scalar, which is correct in slow-roll inflation, but not in more 
general situations such as considered here.}. In this case \eqref{new3} becomes,
\begin{align}\nonumber
\left(\frac{\partial^2 }{\partial N^2}+(3-\epsilon_1)\frac{\partial }{\partial N}\right)\phi_{IR}+{\frac{V'(\phi_{IR})}{H^2}}\\ \label{new4}
+\left[\frac{\partial^2}{\partial N^2}+(3-\epsilon_1)\frac{\partial }{\partial N}{ +}\left({ -\frac{\Delta}{(Ha)^2}}-\frac32\epsilon_2+\frac{1}{2}\epsilon_1\epsilon_2-\frac14\epsilon_2^2-\frac{1}{2}\epsilon_2\epsilon_3\right)\right]\delta\phi_{UV}=0
\,.
\end{align}

Let us now give a more formal definition for $\phi_{IR}$ and $\delta\phi_{UV}$:
\begin{align} \label{eq0.1}
\phi_{IR} (\tau,\textbf{x})&=\int \frac{d^3k}{(2\pi)^{3/2}}\theta(\sigma a(\tau)H(\tau)\!-\!k)e^{-i\textbf{k}\cdot\textbf{x}}\phi_{\textbf{k}}(\tau) \,,
\\ \label{eq0.1.0}
\delta\phi_{UV} (\tau,\textbf{x})&=\int \frac{d^3k}{(2\pi)^{3/2}}\theta(k\!-\!\sigma a(\tau)H(\tau))e^{-i\textbf{k}\cdot\textbf{x}}\phi_{\textbf{k}}(\tau)
\,,
\end{align}
where $\sigma\ll 1$ is a dimensionless cutoff scale and 
$\theta $ is the window function, which for simplicity is taken here to be the Heaviside function. With these definitions we get
\begin{eqnarray}  
\nonumber
\left[\frac{\partial^2}{\partial N^2}+(3-\epsilon_1)\frac{\partial}{\partial N}-\frac{\Delta}{(Ha)^2}\right]\phi_{IR}\!+\!\frac{V'(\phi_{IR})}{H^2}
&\\
\nonumber
=-\int \frac{d^3\textbf{k}}{(2\pi)^{3/2}}e^{-i\textbf{k}\cdot\textbf{x}}
      \left[2\frac{\partial}{\partial N}\theta(k\!-\!\sigma aH)\frac{\partial}{\partial N}\phi_{\textbf{k}}(N)\right]&
\\ 
\nonumber
-\int \frac{d^3\textbf{k}}{(2\pi)^{3/2}}e^{-i\textbf{k}\cdot\textbf{x}}\left[(3-\epsilon_1)
  \left(\frac{\partial}{\partial N}\theta(k\!-\!\sigma aH)\right)\phi_{\textbf{k}}\right]&
\\ 
-\int \frac{d^3\textbf{k}}{(2\pi)^{3/2}}e^{-i\textbf{k}\cdot\textbf{x}}
             \left[\left(\frac{\partial^2}{\partial N^2}\theta(k\!-\!\sigma aH)\right)\phi_{\textbf{k}}\right]&
\,,
\label{eq4}
\end{eqnarray}
where we have already substituted the solution of the second line in Eq.~(\ref{new4})
 for the UV { field}. Defining the stochastic forces $\xi_1(N,\mathbf{x})$ and $\xi_2(N,\mathbf{x})$ as 
\begin{eqnarray} 
\label{eq:5}
\xi_1&=&\sigma aH(1\!-\!\epsilon_1)\int \frac{d^3k}{(2\pi)^{3/2}}\delta(k\!-\!{ \sigma} aH)e^{-i\textbf{k}\cdot\textbf{x}}\phi_{\textbf{k}}(N)\,,
\\ \label{eq:6}
\xi_2&=&\sigma aH(1\!-\!\epsilon_1)\int \frac{d^3k}{(2\pi)^{3/2}}\delta(k\!-\!{ \sigma} aH)e^{-i\textbf{k}\cdot\textbf{x}}
             \frac{\partial\phi_{\textbf{k}}(N)}{\partial N}\,,
\end{eqnarray}
equation~\eqref{eq4} becomes:
\begin{align}
 \label{eq:7}
\left[\frac{\partial^2}{\partial N^2}+(3\!-\!\epsilon_1)\frac{\partial}{\partial N}\right]\phi_{IR}+\frac{V'(\phi_{IR})}{H^2}
=(3\!-\!\epsilon_1)\xi_1+\frac{\partial \xi_1}{\partial N}+\xi_2 { \;\equiv\; \frac{3H}{2\pi}\xi}
\,,
\end{align}
where we made use of,
%
 $\frac{\partial}{\partial N}\theta(k\!-\!\sigma aH)=-\sigma aH(1\!-\!\epsilon_1)\delta(k\!-\!\sigma aH)\,.$
%
Since the linear field { $\phi_{\mathbf{k}}$} is Gaussian, so are the stochastic forces. Thus we have that $\langle\xi_i\rangle=0$ with correlations
\begin{eqnarray} 
\label{eq:8}
\langle\xi_1(N_1,\textbf{x}_1)\xi_1(N_2,\textbf{x}_2)\rangle
   =&\frac{(\sigma a H)^3}{2\pi^2}(1\!-\!\epsilon_1)\left|\varphi(\tau,k)\right|^2_{k=\sigma a H}
       \frac{\sin(\sigma a H r)}{\sigma a H r} \delta (N_1\!-\!N_2) \,,\\ 
\label{eq:9}
\langle\xi_2(N_1,\textbf{x}_1)\xi_2(N_2,\textbf{x}_2)\rangle
   =&\frac{(\sigma a H)^3}{2\pi^2}(1\!-\!\epsilon_1)\left|\frac{\partial\varphi(\tau,k)}{\partial N}\right|^2_{k=\sigma a H}
             \frac{\sin(\sigma a H r)}{\sigma a H r} \delta (N_1\!-\!N_2)\,, 
\\ \label{eq:10}
\langle\xi_1(N_1,\textbf{x}_1)\xi_2(N_2,\textbf{x}_2)\rangle
=&\frac{(\sigma a H)^3}{2\pi^2}(1\!-\!\epsilon_1)\left(\varphi(\tau,k)\frac{\partial\varphi^{*}(\tau,k)}{\partial N}\right)_{k=\sigma a H}
\frac{\sin(\sigma a H r)}{\sigma a H r} \delta (N_1\!-\!N_2)\,,\quad
\end{eqnarray}
where $r=\|\textbf{x}_1\!-\!\textbf{x}_2\|$.~\footnote{The precise spatial dependence of the noise correlators
(given by $\sin(\sigma a H r)/(\sigma a H r)$ in Eqs.~(\ref{eq:8}--\ref{eq:10}))  depends on the 
window function used \cite{Montemayor}. The oscillations in~(\ref{eq:8}--\ref{eq:10}))  
are due to the sharp momentum cutoff imposed by the Heaviside theta function.
Had we used a smooth window function, we would have obtained a function that smoothly interpolates between {\it one}
when $r\ll 1/(\sigma a H)$ and {\it zero} when $r\gg 1/(\sigma a H)$, which can be approximated by the top-hat window function,
$W_{\rm TH}(t,r)=\theta(1-\sigma a Hr)$.
\label{footnote 1}} The correlators are Markovian in the time domain and their decay on super-Hubble scales 
can be modelled by the top-hat window function,
$W_{\rm TH}(t,r)=\theta(1-\sigma a Hr)$, see footnote~\ref{footnote 1}.

Note that Eq. \eqref{eq:7} can be also equivalently found via the Hamiltonian formalism \cite{Grain:2017dqa, Weenink:2011dd} where the momenta and the field satisfy two distinct first order differential equations. Stochastic and Hamiltonian formulations are then equivalent, as long as one keeps track on the separate noise contributions $\xi_1$ and $\xi_2$.

\subsubsection{Regime of validity} \label{Regime}

In this section we will prove that the stochastic approach in USR (and in constant-roll) can only be used at zeroth order in the slow-roll parameters due to the failure of the separate universe approach at leading order in $\epsilon_1$. This differs from the SR case where the separate universe approach is valid at leading order in $\epsilon_1$.

As we have already mentioned, in the separate universe approach, the IR field evolves in a perturbed, but still of a FRW-type, Universe, and its evolution equation follows a Klein-Gordon equation in that separate Universe.  The question is then whether the linearisation of \eqref{eq:7}, once the noises are switched off, reproduces the MS equation. 

We will now expand $\phi_{IR}=\phi_0+\delta\phi_{IR}$ and plug it in the homogeneous part of equation \eqref{eq:7}, obtaining
\begin{eqnarray} 
&&\hskip -2cm
{
2\frac{\delta H}{H}\left[\frac{d^2 \phi_0}{d N^2}+(3\!-\!\epsilon_1)\frac{d \phi_0}{d N}\right]
           -\delta \epsilon_1\frac{\partial\phi_0}{\partial N}
}
\nonumber \\ 
&&\hskip 0cm
{
+\,\frac{\partial^2 \delta\phi_{IR}}{\partial N^2}+ (3\!-\!\epsilon_1)\frac{\partial \delta\phi_{IR}}{\partial N}
+\frac{\delta V'({ \phi_0+\phi_{IR}})}{H^2} +{\cal O}(\delta\phi_{IR}^2)=0
\,.
}
\label{mukfalse}
\end{eqnarray}
Keeping in mind that generally 
we are not in attractor inflation 
-- {\it i.e.} $\partial\phi_{IR}/\partial N$ must be taken as independent of $\phi_{IR}$ --
we have, 
{
\begin{eqnarray} 
\nonumber
 \frac{\delta H}{H} &\simeq& 
\frac{1}{6M_{\rm P}^2}\left[\frac{d \phi_0}{d N}\frac{\partial \delta\phi_{IR}}{\partial N}
+\frac{V'(\phi_0)}{H^2}\delta\phi_{IR}\right]
 = \frac{\sqrt{2\epsilon_1}}{6M_{\rm P}}\left[\frac{\partial \delta\phi_{IR}}{\partial N}
                               -\Big(3\!-\!\epsilon_1\!+\!\frac{\epsilon_2}{2}\Big)\delta\phi_{IR}\right]
 \nonumber
\\ 
\frac{\delta V'}{H^2}&\simeq& \frac{V''(\phi_0)}{H^2}\delta\phi_{IR}
                 = \left[
                     6\epsilon_1-\frac32\epsilon_2-2\epsilon_1^2+\frac52\epsilon_1\epsilon_2-\frac14\epsilon_2^2-\frac12\epsilon_2\epsilon_3
                   \right]\delta\phi_{IR}
 \nonumber
\\ 
\delta \epsilon_1&\simeq& 
\frac{\dot\phi_0\delta\dot\phi_{IR}}{H^2M_{\rm P}^2}-2\epsilon_1\frac{\delta H}{H}
=\frac{\sqrt{2\epsilon_1}}{3M_{\rm P}}\left[(3\!-\!\epsilon_1)\frac{\partial \delta\phi_{IR}}{\partial N}
                               +\Big(3\epsilon_1\!-\!\epsilon_1^2\!+\!\frac{1}{2}\epsilon_1\epsilon_2\Big)\delta\phi_{IR}\right]
\,.\quad
\label{B: delta H}
\end{eqnarray}
}
{ Upon inserting these results into~(\ref{mukfalse}) we get,\footnote{Notice that $\delta\epsilon_1$ is not uniquely specified by stochastic $\Delta N$ formalism. Had we interpreted $\delta \epsilon_1$ as, $\delta \epsilon_1=-\delta\partial_N[\ln(H)] =- \partial_N \delta [\ln(H)]$, we still would not have obtained the correct linear equation (2.38) for $\delta\phi$.}
{\begin{eqnarray} 
\nonumber
&&\hskip -1cm
\frac{\partial^2\delta\phi_{IR}(N)}{\partial N^2}
        +\left(3-\epsilon_1+\frac13\epsilon_1\epsilon_2\right)\frac{\partial\delta\phi_{IR}}{\partial N}
\\
&&+\,\left[-\frac32\epsilon_2+\frac12\epsilon_1\epsilon_2-\frac14\epsilon_2^2-\frac12\epsilon_2\epsilon_3
           +\frac13\epsilon_1^2\epsilon_2-\frac16\epsilon_1\epsilon_2^2\right]\delta\phi_{IR}
                +{\cal O}(\delta\phi_{IR}^2)   =0
                  \,.
\label{B: linear perturbation equation}
\end{eqnarray}
}
{
This equation differs in two important aspects from the equation obeyed by the linearized cosmological perturbations $\delta\phi$, 
which can be inferred from~(\ref{new4}) to read (in the limit when $k\ll aH$),
\begin{equation}
\left[ \frac{\partial^2}{\partial N^2}+(3\!-\!\epsilon_1)\frac{\partial}{\partial N}
   +\left(-\frac32\epsilon_2+\frac12\epsilon_1\epsilon_2-\frac14\epsilon_2^2-\frac12\epsilon_2\epsilon_3
      \right)
\right]\delta\phi(N,\mathbf{x}) +{\cal O}(\delta\phi^2) \simeq  0
\,:
\label{cosmological perturbations: echt}
\end{equation}

\begin{enumerate}
\item[A.]
  While the Hubble damping in the equation for the scalar cosmological perturbation~(\ref{cosmological perturbations: echt})
 is $\gamma = 3-\epsilon_1$, 
there is an additional damping of second order in slow-roll parameters in the stochastic equation of $\delta \gamma_{\rm stoch}=\epsilon_1\epsilon_2/3$ and
\item[B.] While the effective mass term in Eq.~(\ref{cosmological perturbations: echt}) is 
          $\frac{m^2}{H^2}=-\frac32\epsilon_2+\frac12\epsilon_1\epsilon_2-\frac14\epsilon_2^2{ -}\frac12\epsilon_2\epsilon_3$,
the effective mass term in the stochastic equation~(\ref{B: linear perturbation equation})
 has additional terms that are cubic in slow roll parameters, 
 $\frac{\delta m^2_{\rm stoch}}{H^2} =\frac13\epsilon_1^2\epsilon_2-\frac16\epsilon_1\epsilon_2^2$.
\end{enumerate}
The important lesson to take from these observations is that, while in slow roll approximation stochastic $\Delta N$ formalism, 
reproduces correctly the dynamics of the IR quantum field, this is not so in more general cosmological  backgrounds such as inflation 
in non-attractor regime. This disagreement can be traced back to the different constraint structure of the homogeneous (zero) modes and 
inhomogeneous modes of the system gravity plus scalar matter. We postpone a detailed study  of this question for future work.
Needless to say is that a disagreement at the linear order necessarily implies a disagreement at the quantum loop level implying that the
stochastic $\Delta N$ formalism does not solve the question of how to stochasticize the quantum gravity of inflation.

In the following subsections we study in some detail the differences in the predictions of stochastic $\Delta N$ formalism 
in SR and USR regimes. 
}

\bigskip

\paragraph{USR.} In USR where $\epsilon_1^{(USR)} \propto \frac{e^{-6N}}{H^2}$ and hence $\epsilon_2^{(USR)}=-6+2\epsilon_1^{(USR)}$, $\epsilon_3^{(USR)}=2\epsilon_1^{(USR)}$ and
{ Eq.~(\ref{B: linear perturbation equation}) gives},
\begin{align} \label{new5}
\frac{\partial^2 \delta\phi_{IR}}{\partial N^2}
+3\left(1-\epsilon_1^{(USR)}{+\frac29(\epsilon_1^{(USR)})^2}\right)\frac{\partial\delta\phi_{IR}}{\partial N}
  =0
\,,
\end{align} 
or, in conformal time
\be
\left[\frac{\partial^2 }{\partial \tau^2}
          +{\cal H}\Big(\!-2\epsilon_1+\frac{2}{3}\epsilon_1^2\Big)\frac{\partial}{\partial \tau}-\frac{\left(\bar{z}^{(USR)}\right)''}{\bar{z}^{(USR)}}\right](a\delta\phi_{IR})=0\ ,
\label{new new}
\ee
where
\be \label{new6} \quad \frac{\left(\bar{z}^{(USR)}\right)''}{\bar{z}^{(USR)}}={\cal H}^2\left(2-\epsilon_1^{(USR)}
+\frac23(\epsilon_1^{(USR)})^2\right)
\,.
\ee
As we clearly see \eqref{new6}, calculated via the separate universe approach, differs from \eqref{zzUSR}, calculated via the MS equation, at leading (linear) order in $\epsilon_1$. In addition,~(\ref{new new}) contains an
anti-damping term that is linear in slow roll. 
Therefore the separate universe formalism, and thus the stochastic formalism, fails at leading order in slow-roll to reproduce the correct evolution of the linear fluctuations obtained from the quantum field theory approach. 

We would like here to briefly mention the constant-roll case (CR). There $\frac{\ddot{\phi}}{3H\dot{\phi}}=\frac{\epsilon_2-2\epsilon_1}{6}=\textit{constant}$. Constant roll is supported only by a specific class of inflationary potentials \cite{Motohashi:2014ppa}. In some of these potentials $\epsilon_1$ decreases and $\epsilon_2$ reaches a non-negligible constant which can be any value. Because of this, again, the stochastic formalism fails beyond the zeroth order in $\epsilon_1$. In addition, to keep $\frac{\delta m_{stoch}^2}{H^2}\ll 1$ so to be able to use the stochastic formalism \textit{at least} at zeroth order in $\epsilon_1$, we want $\delta m_{stoch}$ to be small so we get the constraint  (at leading order in $\epsilon_1\ll1$)
\be
\epsilon_2^2\ll\frac{6}{\epsilon_1}\ .\label{conep}
\ee  
For typical values of $\epsilon_2$ that are found in models of inflation related to PBHs formations, \eqref{conep} implies that during the CR or USR phase $\epsilon_1\ll 10^{-1}$.

\paragraph{SR} The same comparison can be now easily done in SR. In this case one can see that if we keep ourselves to first order in $\epsilon_1$ we get from \eqref{B: linear perturbation equation}, rewritten in conformal time,
\begin{align} \label{SRlin}
\left[\frac{\partial^2 }{\partial \tau^2}-\frac{\left(\bar{z}^{(SR)}\right)''}{\bar{z}^{(SR)}}\right](a\delta\phi_{IR})=0
\,, \quad \text{where} \quad \frac{\left(\bar{z}^{(SR)}\right)''}{\bar{z}^{(SR)}}={\cal H}^2\left(2-\epsilon_1^{(SR)}
{+\frac32\epsilon_2}+{\cal O}((\epsilon_1^{(SR)})^2)\right)
\,,
\end{align}
which, because in SR $\epsilon_2\ll 1$, matches the linear analysis at leading  { (linear) order in slow-roll but --
 as one can easily show -- does not match at second order in slow roll.

In summary, we have shown that the stochastic $\Delta N$ formalism reproduces correctly the linear dynamics of scalar cosmological perturbations
to linear order in slow-roll parameters (albeit it fails at higher order) in SR, but it fails to reproduce correctly the corresponding dynamics at leading linear order in slow-roll parameters in USR and CR regimes. The origin of that failure can be traced back to the fact that stochastic $\Delta N$ formalism 
does not correctly incorporate the gravitational constraints. That this is so can be seen {\it e.g.} from the fact that 
in stochastic $\Delta N$ formalism the effective mass of $\delta\phi_{IR}$ field vanishes in USR in which $V''=0$ 
(see Eq.~(\ref{new5})). However, as one can easily convince oneself from
Eq.~(\ref{cosmological perturbations: echt}), the true scalar cosmological perturbations have a non-vanishing effective mass
 even when $V''(\phi)=0$  (the mass is given by $m^2= -2\epsilon_1H^2(3-\epsilon_1)$), which is generated by the gravitational constraint.

\subsubsection{Final version of the Langevin equation for USR}

Having found out that the stochastic formalism during USR is only valid at zeroth order in $\epsilon_1$, by using the solution \eqref{SOL} at this order we obtain
\begin{align} \label{eq:32}
\langle\xi_1(N_1)\xi_1(N_2)\rangle\simeq \left(\frac{H}{2\pi}\right)^2\delta(N_1-N_2)\,, \\ \label{eq:33}
\langle\xi_2(N_1)\xi_2(N_2)\rangle\simeq 0\,,  \\ \label{eq:34}
\langle\xi_2(N_1)\xi_1(N_2)\rangle\simeq 0\,.
\end{align}
Finally, by using \eqref{eq:32}-\eqref{eq:34} in \eqref{eq:7} we have
\begin{align} \label{eq:langevin eq}
\frac{\partial^2\phi_{IR}}{\partial N^2}+3\frac{\partial\phi_{IR}}{\partial N}=\frac{3H}{2\pi}\xi(N)\,,
\end{align}
where we have approximated, $\xi_1\approx \left(\frac{H}{2\pi}\right)\xi$, $\xi_2\simeq 0$
 so that $\langle \xi(N)\xi(N')\rangle\approx\delta(N-N')$. 
 
 We immediately see that \eqref{eq:langevin eq} is {\it linear} in $\phi_{IR}$. The reason is that all the non-linearities are hidden in the slow-roll parameters that cannot be evaluated exactly at this level. Nevertheless, precisely as in the SR case, they only provide negligible corrections to the power spectrum that, by consistency, should match the power spectrum calculated in the quantum field theory approach to the linear theory. This is what we are going to show next. 

\section{Stochastic power spectrum in USR}

Defining $V_N=e^{3N}\frac{\partial\phi_{IR}}{\partial N}\,,$ we can write \eqref{eq:langevin eq} as a couple of first order stochastic differential equations:
\begin{align} \label{eq:36}
V_N=e^{3N}\frac{\partial\phi_{IR}}{\partial N}\,,
\\ \label{eq:37}
\frac{\partial V_N}{\partial N}=\frac{3H}{2\pi}e^{3N}\xi(N)\,.
\end{align}
Integrating \eqref{eq:37} we get
\begin{align} \label{eq:38}
V_N(N)-V_N(0)=\int_0^N\frac{3H}{2\pi}e^{3N'}\xi(N')dN'\,,
\end{align}
where $V_N(0)$ is the classical value of the velocity of the field at the beginning of the USR phase, {\it i.e.}
{$V_N(0)=\frac{\partial \phi_{IR}(0)}{\partial N}=\frac{\dot{\phi}_{IR}(0)}{H(0)}=\sqrt{2\epsilon_1(0)}M_{\rm P}\,,$
 so that:} 
\begin{align} \label{eq:39}
V_N(N)=\int_0^N\frac{3H}{2\pi}e^{3N'}\xi(N')dN'+\sqrt{2\epsilon_1(0)}M_{\rm P}\,.
\end{align}
From \eqref{eq:36} we have $\frac{\partial\phi_{IR}}{\partial N}=e^{-3N}V_N(N)$, setting $\phi_{IR}(0)=0$ (the theory is shift invariant) we obtain
\begin{align} \label{eq;40}
\phi_{IR}(N)=\int_0^N e^{-3N''}V_N(N'')dN''=\int_0^N e^{-3N''}dN''\left[\int_0^{N''}\frac{3H}{2\pi}e^{3N'}\xi(N')dN'
+{\sqrt{2\epsilon_1(0)}M_{\rm P}}\right]
\,.
\end{align}
As we are interested {in} the fluctuations around the homogeneous mean value, 
{\it i.e.} the power spectrum of the fluctuations, we will consider the following quantity:
\begin{align} \label{eq:42}
\phi_{IR}(N)-\langle\phi_{IR}(N)\rangle=\delta\phi_{IR}(N)=\int_0^N e^{-3N''}dN''\int_0^{N''}\frac{3H}{2\pi}e^{3N'}\xi(N')dN'\,.
\end{align}
The power spectrum of the scalar perturbations is then proportional to the two-point correlation function $\langle\delta\phi_{IR}(N)\delta\phi_{IR}(\bar{N})\rangle_{N\rightarrow\bar{N}}$. Using the statistical properties of the noise $\xi$ we get
\begin{align} \label{eq:450}
\langle\delta\phi_{IR}(N)\delta\phi_{IR}(N)\rangle=\frac{H^2}{4\pi^2}\left(N-\frac{1}{2}-\frac{e^{-6N}}{6}+2\frac{e^{-3 N}}{3 }\right)\,.
\end{align}
Because we are only considering the stochastic formalism at leading order in slow-roll parameters, we need to consistently neglect the decaying modes that are proportional to powers of $\frac{\epsilon_1}{\epsilon_1^*}$, where $\epsilon_1^*$ is the slow-roll parameter at crossing horizon. Then we finally get 
\begin{align} \label{eq:45}
\langle\delta\phi_{IR}(N)\delta\phi_{IR}(N)\rangle=\frac{H^2}{4\pi^2}\left(N-\frac{1}{2}\right)\,.
\end{align}
The linear growth in $N$ can be easily understood. Equations \eqref{eq:45}
gives the two point function in real space, which grows as $N=\ln(a)$ 
since -- as time grows -- more and more modes cross the Hubble radius, thereby explaining the $\ln(a)$ growth in real space.

\subsection{Comparison with the linear analysis}
The power spectrum ${\cal P}_{\chi}$ of a field $\chi$ is defined as
\begin{align}\label{powdef}
\langle\chi^2\rangle=\int\frac{dk}{k}{\cal P}_{\chi}
\,.
\end{align}
In USR, the power spectrum calculated by the use of linear analysis is, in Fourier space, \cite{sasaki0}
\begin{align} \label{POW}
{\cal P_{\delta\phi}}=\frac{H^2}{4\pi^2}\left(1+{\cal O}\left(\epsilon_1^{(USR)}\right)\right)\ .
\end{align}
The two point correlation function \eqref{eq:45} is given in real space and strictly speaking for infinitely long wavelength, therefore an anti-Fourier transformation of \eqref{eq:45} would not make sense. 

Nevertheless, we can make use the fact that ${\cal P}_{\delta\phi}$ is dominated by the constant mode in USR already after one e-foldings. In that case, ${\cal P}$ only depends from the value it acquires at {the Hubble crossing}. Then, since $d \ln k =Hdt=dN$, we can calculate the power spectrum from the definition \eqref{powdef} by taking its derivative with respect to $N$ \cite{Kunze}. We then immediately obtain
\begin{align} \label{finalpower}
{\cal P_{\delta\phi}}\simeq\frac{d}{dN}\langle\delta\phi_{IR} \delta\phi_{IR}\rangle\simeq\frac{H^2}{4\pi^2}\ .
\end{align}
Since the quantum kick is only active after one e-fold \cite{Kunze} and at the same time the constant mode dominates 
after roughly one e-fold, the power {spectra} \eqref{POW} and \eqref{finalpower} are, within our approximations, 
the same. This allows us to conclude that stochastic inflation has no significant effects during the USR regime of inflation. Note that, even if we had, somehow inconsistently considered the exponentially decaying terms from \eqref{eq:450}, these would be exactly matched with the ``decaying" modes appearing in the linear solution of the MS equation, as it should.

Finally then, given that the stochastic power spectrum of $\delta\phi$ matches the one of the linear analysis, we are now allowed to use the linear gauge relation $\zeta=\frac{a\delta\phi}{z}$ that relates the curvature perturbations $\zeta$
{ to the scalar field perturbation} $\delta\phi$. Then the power spectrum of the curvature perturbation generated by quantum diffusion effects at linear level reads, 
\begin{align}\label{powerzeta}
{\cal P_{\zeta}}\simeq\frac{H^2}{{ 8\pi^2\epsilon_1 M_{\rm P}^2}}\ .
\end{align}
such that -- as expected -- it reproduces correctly the power spectrum of the canonically quantised curvature perturbation, but no corrections to it. 

In this section we have found that the stochastic evolution at zeroth order in the slow-roll parameters exactly reproduces what we expect from the quantum field theory result. This is what one should expects as for a flat potential and at zeroth order in the slow-roll parameters, there are no self-interactions terms of the scalar field. Of course, the self-interactions are important beyond the linear lever, but that is beyond the scope of our paper.

\section{Conclusions}

In this paper we have shown that the separate universe approach, where perturbations are supposed to follow a local Klein-Gordon equations in a separate FRW universe generically fails at leading (linear) order in the slow-roll parameters and at any orders whenever $\epsilon_2^2\gtrsim 6/\epsilon_1$. This immediately implies the generic failure of the stochastic approach. The failure of the standard stochastic $\Delta N$ formalism does not imply that it cannot be correctly re-formulated. In fact, the general idea that spatial gradients can be neglected on long wavelengths it is clearly correct for any local gravitational theory. Thus, one could use it as a starting point for a successful construction of a novel stochastic $\Delta N$ formalism.

In USR $\epsilon_2^2\ll 6/\epsilon_1$ and thus the stochastic approach can be used at zeroth order in slow-roll parameters. There, because the potential is flat, the scalar field is effectively free in any local universe immediately implying that the higher order stochastic effects should be small. Indeed,
by directly calculating the two-point correlation function of the inflaton in the stochastic formalism in USR, we show that this, up to negligible slow-roll corrections, coincides with the power spectrum calculated by linear perturbation theory. This proves that, in USR and at the zeroth order in slow-roll parameters, the entire leading order contribution to the power spectrum 
of the curvature perturbation is captured by the free quantum diffusion of the inflation. This result can then be used 
to obtain a rough estimate for the formation of PBHs \cite{tomi}. 
Because the jump in the spectral index of the mode function at the SR-USR transition is ${\cal O}(\epsilon_1)$, 
we do not expect the result to change significantly when the SR-USR matching is properly included in the analysis\footnote{Note that the conclusion may be different for a test field in an ultra-slow roll background, see e.g. \cite{vennin}.
The SR-USR mode matching is expected to generate a mixing between the growing and decaying modes that is suppressed as $\epsilon_1$, and hence can be neglected at the leading (zeroth) other analysis in slow roll. If one is interested in the effects that are linear in slow roll parameters however, the mode matching should be included. This cannot unfortunately be done in the current stochastic approach. We intend to address this 
 interesting question in the context of quantum field theory of inflation elsewhere. }
 
As we have already commented in the introduction, our result \eqref{powerzeta} differs from the equivalent one found by \cite{bellido}. The mistake in \cite{bellido} is rooted in the use of the approach of \cite{Kunze} to translate the real space $\langle\zeta^2\rangle$, calculated with stochastic methods, to the Fourier space. Although this is correct in SR, where the super-horizon $\zeta$ is constant, in USR $\zeta$ is dominated by the growing mode and therefore the 
approach advocated in~\cite{Kunze} cannot be used.

A separate question that has been asked in \cite{riotto2} was whether subleading 
quantum diffusion effects might change the predictions of the PBHs abundances calculated in USR. Given the fact that at the zeroth order in the slow-roll parameters the predictions from stochastic inflation precisely match the ones from the quantum field theory of linear perturbations, it should be obvious that, 
free quantum diffusion should suffice to get a rough estimate of the PBH abundance. This point of view 
is not shared by \cite{riotto2}. There, the authors have promoted the power spectrum of the curvature perturbation
${\cal P}$, which is in the linear theory an expectation value of the two point function, 
to a stochastic variable. Thus, in their case, any derived quantity of it must be 
re-mediated. We disagree on this precisely because $\cal P$ is {\it already} a mediated quantity 
depending only on the initial conditions for the stochastic evolution. 

To conclude, any corrections to the free quantum diffusion in the calculation of PBHs abundances reside in the slow-roll corrections to the stochastic inflation, which are bound to be small as they are slow-roll suppressed. 
However, nonlinear (loop) quantum gravitational effects may still be significant~\cite{vennin,Boran:2016bla}. 
In order to calculate them, one would then 
have to estimate their production either by performing a loop QFT calculation or within a suitably improved stochastic 
framework that properly takes account of the gravitational constraints. The principle obstacles on how to construct such a framework are discussed in \cite{Tsamis:2005hd,Miao:2018bol}}. 

\appendix
\section{Derivation of the noises at leading order in slow-roll}

Since in SR inflation the separate Universe approach is valid at leading order in slow-roll parameters, we provide here the stochastic forces at this order.

The procedure to find them is exactly the same as done before. We start from \eqref{eq:8}-\eqref{eq:10} but now using the mode function \eqref{SOL} up to first order in $\epsilon_1$
\begin{align}
\varphi(\tau,k)=\frac{1}{a}\sqrt{\frac{-\pi\tau}{4}}H_\nu^{(1)}(-k\tau)
\,, \quad \text{with} \quad \nu=\frac{3}{2}+\epsilon_1\ .
\end{align} 
Then, we finally get
\begin{align} \label{SRnoise1}
\langle\xi_1(N_1)\xi_1(N_2)\rangle\simeq \left[1+2\left(\psi^{(0)}\left(\frac{3}{2}\right)-\log(\sigma/2)-\frac{1}{2}\right)\epsilon_1\right]\left(\frac{H}{2\pi}\right)^2\delta(N_1-N_2)\,, \\ \label{SRnoise2}
\langle\xi_2(N_1)\xi_2(N_2)\rangle\simeq 0\,,  \\ \label{SRnoise3}
\langle\xi_2(N_1)\xi_1(N_2)\rangle\simeq \epsilon_1\left(\frac{H}{2\pi}\right)^2\delta(N_1-N_2)\,,
\end{align}
where $\psi^{(0)}$ is the digamma function.

\acknowledgments

CG thanks Toni Riotto for discussions. CG is supported by the Ramon y Cajal program and partially supported by the Unidad de Excelencia Mar\'ia de Maeztu Grant No. MDM-2014-0369 and by the national FPA2013-46570-C2-2-P and FPA2016-76005-C2-2-P grants. DCM is supported by the national ``Beca de colaboracion en departamentos''. 
T.P. acknowledges support from the D-ITP consortium, 
a program of the NWO that is funded by the Dutch Ministry of 
Education, Culture and Science (OCW) and a support by a research programme of the Foundation for Fundamental Research on Matter (FOM), which is part of the Netherlands Organisation for Scientific Research (NWO).


\begin{thebibliography}{99}
\bibitem{pbhs}
S.~Bird, I.~Cholis, J.~B.~Munoz, Y.~Ali-Haimoud, M.~Kamionkowski, E.~D.~Kovetz, A.~Raccanelli and A.~G.~Riess,
  ``Did LIGO detect dark matter?,''
  Phys.\ Rev.\ Lett.\  {\bf 116} (2016) no.20,  201301
  doi:10.1103/PhysRevLett.116.201301
  [arXiv:1603.00464 [astro-ph.CO]]; Y.~Ali-Haimoud and M.~Kamionkowski,
  ``Cosmic microwave background limits on accreting primordial black holes,''
  Phys.\ Rev.\ D {\bf 95} (2017) no.4,  043534
  doi:10.1103/PhysRevD.95.043534
  [arXiv:1612.05644 [astro-ph.CO]].
\bibitem{Kinney:2005vj}
  W.~H.~Kinney,
  ``Horizon crossing and inflation with large eta,''
  Phys.\ Rev.\ D {\bf 72} (2005) 023515
  doi:10.1103/PhysRevD.72.023515
  [gr-qc/0503017].
\bibitem{Martin:2012pe}
  J.~Martin, H.~Motohashi and T.~Suyama,
  ``Ultra Slow-Roll Inflation and the non-Gaussianity Consistency Relation,''
  Phys.\ Rev.\ D {\bf 87} (2013) no.2,  023514
  doi:10.1103/PhysRevD.87.023514
  [arXiv:1211.0083 [astro-ph.CO]].
\bibitem{tomi}
C.~Germani and T.~Prokopec,
  ``On primordial black holes from an inflection point,''
  Phys.\ Dark Univ.\  {\bf 18} (2017) 6
  doi:10.1016/j.dark.2017.09.001
  [arXiv:1706.04226 [astro-ph.CO]].
\bibitem{musco}
C.~Germani and I.~Musco,
  ``The abundance of primordial black holes depends on the shape of the inflationary power spectrum,''
  arXiv:1805.04087 [astro-ph.CO].
  \bibitem{toni}
  G.~Franciolini, A.~Kehagias, S.~Matarrese and A.~Riotto,
  ``Primordial Black Holes from Inflation and non-Gaussianity,''
  JCAP {\bf 1803} (2018) no.03,  016
  doi:10.1088/1475-7516/2018/03/016
  [arXiv:1801.09415 [astro-ph.CO]].
\bibitem{sasaki}
Y.~F.~Cai, X.~Chen, M.~H.~Namjoo, M.~Sasaki, D.~G.~Wang and Z.~Wang,
  ``Revisiting non-Gaussianity from non-attractor inflation models,''
  JCAP {\bf 1805} (2018) no.05,  012
  doi:10.1088/1475-7516/2018/05/012
  [arXiv:1712.09998 [astro-ph.CO]].
\bibitem{Starobinsky3}
  A.~A.~Starobinsky,
  ``Stochastic De Sitter (inflationary) Stage In The Early Universe,''
  Lect.\ Notes Phys.\  {\bf 246} (1986) 107.
  doi:10.1007/3-540-16452-9\_6

\bibitem{vennin}
C.~Pattison, V.~Vennin, H.~Assadullahi and D.~Wands,
  ``Quantum diffusion during inflation and primordial black holes,''
  JCAP {\bf 1710} (2017) no.10,  046
  doi:10.1088/1475-7516/2017/10/046
  [arXiv:1707.00537 [hep-th]].
\bibitem{riotto2}
M.~Biagetti, G.~Franciolini, A.~Kehagias and A.~Riotto,
  ``Primordial Black Holes from Inflation and Quantum Diffusion,''
  arXiv:1804.07124 [astro-ph.CO].
  \bibitem{bellido}
  J.~M.~Ezquiaga and J.~García-Bellido,
  ``Quantum diffusion beyond slow-roll: implications for primordial black-hole production,''
  arXiv:1805.06731 [astro-ph.CO].
\bibitem{LindeInflation}
  A.~D.~Linde,
  ``A New Inflationary Universe Scenario: A Possible Solution of the Horizon, Flatness, Homogeneity, Isotropy and Primordial Monopole Problems,''
  Phys.\ Lett.\  {\bf 108B} (1982) 389.
  doi:10.1016/0370-2693(82)91219-9
\bibitem{Liddle}
  A.~R.~Liddle, P.~Parsons and J.~D.~Barrow,
  ``Formalizing the slow roll approximation in inflation,''
  Phys.\ Rev.\ D {\bf 50} (1994) 7222
  doi:10.1103/PhysRevD.50.7222
  [astro-ph/9408015].
\bibitem{mukhanov}
  V.~F.~Mukhanov, H.~A.~Feldman and R.~H.~Brandenberger,
  ``Theory of cosmological perturbations. Part 1. Classical perturbations. Part 2. Quantum theory of perturbations. Part 3. Extensions,''
  Phys.\ Rept.\  {\bf 215} (1992) 203.
  doi:10.1016/0370-1573(92)90044-Z
\bibitem{Vilenkin}
  A.~Vilenkin,
  ``Quantum Fluctuations in the New Inflationary Universe,''
  Nucl.\ Phys.\ B {\bf 226} (1983) 527.
  doi:10.1016/0550-3213(83)90208-0
\bibitem{Salopek}
  D.~S.~Salopek and J.~R.~Bond,
  ``Nonlinear evolution of long wavelength metric fluctuations in inflationary models,''
  Phys.\ Rev.\ D {\bf 42} (1990) 3936.
  doi:10.1103/PhysRevD.42.3936
\bibitem{Kandrup}
  H.~E.~Kandrup,
  ``Stochastic Inflation As A Time Dependent Random Walk,''
  Phys.\ Rev.\ D {\bf 39} (1989) 2245.
  doi:10.1103/PhysRevD.39.2245

\bibitem{Salopek2}
  D.~S.~Salopek and J.~R.~Bond,
  ``Stochastic inflation and nonlinear gravity,''
  Phys.\ Rev.\ D {\bf 43} (1991) 1005.
  doi:10.1103/PhysRevD.43.1005


\bibitem{Mukhanov:1990me}
  V.~F.~Mukhanov, H.~A.~Feldman and R.~H.~Brandenberger,
  ``Theory of cosmological perturbations. Part 1. Classical perturbations. Part 2. Quantum theory of perturbations. Part 3. Extensions,''
  Phys.\ Rept.\  {\bf 215} (1992) 203.
  doi:10.1016/0370-1573(92)90044-Z


\bibitem{Wands}
  D.~Wands, K.~A.~Malik, D.~H.~Lyth and A.~R.~Liddle,
  ``A New approach to the evolution of cosmological perturbations on large scales,''
  Phys.\ Rev.\ D {\bf 62} (2000) 043527
  doi:10.1103/PhysRevD.62.043527
  [astro-ph/0003278].

\bibitem{Vennin1}
  V.~Vennin and A.~A.~Starobinsky,
  ``Correlation Functions in Stochastic Inflation,''
  Eur.\ Phys.\ J.\ C {\bf 75} (2015) 413
  doi:10.1140/epjc/s10052-015-3643-y
  [arXiv:1506.04732 [hep-th]].

\bibitem{Finelli:2008zg}
 F.~Finelli, G.~Marozzi, A.~A.~Starobinsky, G.~P.~Vacca and G.~Venturi,
Phys.\ Rev.\ D {\bf 79} (2009) 044007
doi:10.1103/PhysRevD.79.044007
[arXiv:0808.1786 [hep-th]].

\bibitem{Montemayor}
  H.~Casini, R.~Montemayor and P.~Sisterna,
  ``Stochastic approach to inflation. 2. Classicality, coarse graining and noises,''
  Phys.\ Rev.\ D {\bf 59} (1999) 063512
  doi:10.1103/PhysRevD.59.063512
  [gr-qc/9811083].

\bibitem{Grain:2017dqa}
J.~Grain and V.~Vennin,
JCAP {\bf 1705} (2017) no.05,  045
doi:10.1088/1475-7516/2017/05/045
[arXiv:1703.00447 [gr-qc]].


\bibitem{Weenink:2011dd}
J.~Weenink and T.~Prokopec,
arXiv:1108.3994 [gr-qc].


\bibitem{Motohashi:2014ppa}
H.~Motohashi, A.~A.~Starobinsky and J.~Yokoyama,
JCAP {\bf 1509} (2015) 018
doi:10.1088/1475-7516/2015/09/018
[arXiv:1411.5021 [astro-ph.CO]].


\bibitem{sasaki0}
M.~H.~Namjoo, H.~Firouzjahi and M.~Sasaki,
  ``Violation of non-Gaussianity consistency relation in a single field inflationary model,''
  EPL {\bf 101} (2013) no.3,  39001
  doi:10.1209/0295-5075/101/39001
  [arXiv:1210.3692 [astro-ph.CO]].

\bibitem{Kunze}
  K.~E.~Kunze,
  ``Perturbations in stochastic inflation,''
  JCAP {\bf 0607} (2006) 014
  doi:10.1088/1475-7516/2006/07/014
  [astro-ph/0603575].


\bibitem{Boran:2016bla}
  S.~Boran and E.~O.~Kahya,
  ``Loop corrections to primordial non-Gaussianity,''
  Phys.\ Rev.\ D {\bf 97} (2018) no.4,  043507
  doi:10.1103/PhysRevD.97.043507
  [arXiv:1601.01106 [astro-ph.CO]].

\bibitem{Tsamis:2005hd}
  N.~C.~Tsamis and R.~P.~Woodard,
  ``Stochastic quantum gravitational inflation,''
  Nucl.\ Phys.\ B {\bf 724} (2005) 295
  doi:10.1016/j.nuclphysb.2005.06.031
  [gr-qc/0505115].

\bibitem{Miao:2018bol}
  S.~P.~Miao, T.~Prokopec and R.~P.~Woodard,
  ``The Graviton Tail almost Completely Wags the Dog,''
  arXiv:1806.00742 [gr-qc].


\end{thebibliography}
\end{document}